\documentstyle[12pt]{article}
\newlength{\pubnumber} 

\catcode`\@=11
\@addtoreset{equation}{section}

\def\section{\@startsection{section}{1}{\z@}{3.5ex plus 1ex minus .2ex}
 {2.3ex plus .2ex}{\large\bf}}
\def\subsection{\@startsection{subsection}{2}{\z@}{2.3ex plus .2ex}
 {2.3ex plus .2ex}{\bf}}

\def\NPB#1#2#3{{\it Nucl.\ Phys.}\/ {\bf B#1} (19#2) #3}

\def\PRD#1#2#3{{\it Phys.\ Rev.}\/ {\bf D#1} (19#2) #3}

\begin{document}

\begin{titlepage}
\samepage{
\setcounter{page}{1}
\rightline{IASSNS-HEP-96/64}
\rightline{\tt hep-ph/9606467}
\rightline{Published: {\it Nucl.\ Phys.}\/ {\bf B488} (1997) 141}
\rightline{September 1996}
\vfill
\begin{center}
 {\Large \bf
    New Constraints on SO(10) Model-Building\\
    from String Theory\\}
\vfill
\smallskip
 {\large Keith R. Dienes\footnote{
   E-mail address: dienes@sns.ias.edu}\\}
\vspace{.10in}
 {\it  School of Natural Sciences, Institute for Advanced Study\\
  Olden Lane, Princeton, N.J.~~08540~ USA\\}
\end{center}
\vfill
\begin{abstract}
  {\rm
    A recent analysis of the methods by which higher-level gauge
    symmetries are realized in string theory has made it possible
    to systematically obtain new constraints on string GUT
    model-building.  In this paper, we perform such a study for the
    case of $SO(10)$ free-field string GUT models, and find a number
    of significant new results.
    First, we show that within the conventional string models
    using the so-called ``diagonal'' embeddings,
   all representations larger than the {\bf 16}\/
    must always transform as singlets under all gauge symmetries beyond
     $SO(10)$.
    This includes the ${\bf 45}$ and ${\bf 54}$ representations, and holds
    regardless of the affine level at which $SO(10)$ is realized.
     Furthermore, we show that
    such constructions can never give rise to
     the ${\bf 120}$, ${\bf 126}$, or ${\bf 144}$ representations
     of $SO(10)$ --- again regardless of the affine level.
     We also perform a similar analysis for
    $E_6$, and find that the
    adjoint ${\bf 78}$ representation must always transform as
    a singlet under all gauge symmetries beyond $E_6$;  moreover, all
    representations beyond the ${\bf 78}$ representation are prohibited.
  Taken together, these results thus severely limit the types
    of phenomenologically realistic field-theoretic $SO(10)$ and $E_6$
     models that can be
    obtained using such string constructions.
   Finally, we also explore the possibility of using
     non-diagonal free-field embeddings,
   and show that in the $SO(10)$ case
     such embeddings
     may have improved phenomenological prospects.}
\end{abstract}
\medskip
   }
\end{titlepage}

\setcounter{footnote}{0}

\def\beq{\begin{equation}}
\def\eeq{\end{equation}}
\def\beqn{\begin{eqnarray}}
\def\eeqn{\end{eqnarray}}
\def\bbox#1{\parbox[t]{1.20 in}{{#1}}}
\def\bboxb#1{\parbox[t]{1.00 in}{{#1}}}
\def\bbreak{\vfill\break}
\def\rep#1{{\bf {#1}}}
\hyphenation{su-per-sym-met-ric non-su-per-sym-met-ric}
\hyphenation{space-time-super-sym-met-ric}
\hyphenation{mod-u-lar mod-u-lar--in-var-i-ant}


\setcounter{footnote}{0}
\setcounter{page}{2}
\section{Introduction}

One of the more popular ways of understanding the apparent unification
of gauge couplings within the minimal supersymmetric standard model (MSSM)
has been the assumption of an unbroken unified gauge symmetry
above the scale of unification.  Such a supersymmetric grand-unified
theory (GUT) could also in principle explain the quantum numbers of the
observed fermion spectrum  of the Standard Model, and may ultimately explain
the origins of fermion mass.  Among the possible GUT models, those utilizing
an $SO(10)$ gauge symmetry are particularly compelling \cite{classics}, since
$SO(10)$ is
the smallest simple Lie group for which a single anomaly-free irreducible
representation
can accommodate the entire Standard Model fermion content of each generation.
Unification via $SO(10)$ can also provide a natural doublet-triplet
splitting mechanism \cite{doublettriplet},
and can accommodate small neutrino masses via the see-saw mechanism
\cite{seesaw}.
There has therefore been considerable interest in $SO(10)$ GUT model-building
in recent years, and many different proposals currently exist
\cite{neutrals,bads,goods}.

One way of constraining the set of possible $SO(10)$ unification models is to
require that they can be realized as the low-energy limits of heterotic
strings.
Indeed, in this way, it is hoped that the entire $SO(10)$ framework can be
joined with gravity in a consistent string model.  For this reason there has
been substantial effort in constructing such string GUT models
\cite{stringGUTs,AFIU,erler,KT},
and a review can be found in Ref.~\cite{review}.
For a long time, the primary difficulty had been the construction of
$SO(10)$ models with three generations.
Recently, however, there have appeared several promising
three-generation $SO(10)$ and $E_6$ string GUT models \cite{KT}.
Unfortunately, much of the analysis of these models remains to be done,
and in particular these models have not yet been shown to be phenomenologically
realistic in other aspects (such as fermion masses, proton lifetime,
and so forth).
Work along these lines is in progress \cite{KT}.
Nevertheless, by studying the {\it generic}\/ model-independent
properties that follow from the string consistency constraints themselves,
valuable guidance for low-energy $SO(10)$ model-building can still be obtained.
This is the primary goal of the present paper.

Certain types of model-independent information are well-known.
For example, in heterotic string models, gauge symmetries are typically
realized
as affine Lie algebras (also called Ka\v{c}-Moody algebras) with a given
affine level.  Thus, simply by studying the unitary representation theory
and central charges of $SO(10)$ affine Lie algebras at various levels,
one can immediately deduce a number of important constraints.
For example, since the central charge of $SO(10)$ at level $k$ is given
by $c=45k/(k+8)$, the perturbative heterotic-string central-charge
constraint $c\leq 22$ implies that one can never realize $SO(10)$ at
levels $k\geq 8$.  Thus, only the cases of $SO(10)$ with levels $1\leq k \leq
7$
need be considered.  This in turn limits the allowed
$SO(10)$ representations that can potentially appear in the massless
heterotic-string spectrum, and
a list of such representations for affine levels $1\leq k\leq 4$
appears in Table~\ref{tableone}.

 From this table, a number of additional constraints can be determined.
First, it is immediately apparent that
the adjoint $\rep{45}$ representation
is possible only in string models in which the $SO(10)$ gauge symmetry
is realized at levels $k\geq 2$.
Since such adjoint representations are required for the GUT Higgs scalar
(in order to break the GUT gauge symmetry down to that of the Standard Model),
we see that any consistent $SO(10)$ string GUT model must in
fact realize the $SO(10)$ gauge symmetry at $k\geq 2$.
Furthermore, if one wishes to obtain a $\rep{120}$ or $\rep{144}$
representation
(as is sometimes used for proper light quark/lepton mass ratios as well
as for GUT symmetry breaking),
one is restricted to levels $k\geq 3$;
likewise, in order to achieve a $\rep{126}$ representation
(as is often used for a heavy Majorana right-handed neutrino
mass, as well as for the Georgi-Jarlskog factor of three in the light
quark/lepton mass ratio \cite{GJ} and for GUT symmetry breaking with
automatic $R$-parity conservation),
it turns out that one would
require $SO(10)$ at levels $k\geq 5$.

\begin{table}[t]
\centerline{
\begin{tabular}{l|l|l|l}
{}~~~$k=1$ & ~~~$k=2$ & ~~~$k=3$ & ~~~$k=4$ \\
\hline
\hline
              (\rep{10},~1/2) &
              (\rep{10},~9/20) &
             (\rep{10},~9/22) &
             (\rep{10},~3/8) \\
             (\rep{16},~5/8) &
             (\rep{16},~9/16) &
             (\rep{16},~45/88) &
             (\rep{16},~15/32) \\
            ~ &
             (\rep{45},~4/5) &
             (\rep{45},~8/11) &
             (\rep{45},~2/3) \\
            ~ &
             (\rep{54},~1) &
             (\rep{54},~10/11) &
             (\rep{54},~5/6) \\
           ~ &
           ~ &
             (\rep{120},~21/22) &
             (\rep{120},~7/8) \\
           ~ &
           ~ &
             (\rep{144},~85/88) &
             (\rep{144},~85/96) \\
           ~ &
           ~ &
           ~ &
             (\rep{210},~1) \\
\hline
\end{tabular}
  }
\caption{Unitary, potentially massless representations of $SO(10)$
realized at affine levels $k=1,2,3,4$.
Each representation $R$ is listed as $({\bf dim\,R},h_{R})$
where $h_R$ is its conformal dimension.
Singlets and conjugate representations are not explicitly
written, but understood.}
\label{tableone}
\end{table}

Further model-independent constraints can also
be gained by studying the {\it conformal dimensions}\/ $h_R$ of the
allowed representations.  In string theory, the conformal dimension
of a given state describes its spacetime mass (in Planck-mass units),
and is related to the number of string excitations needed to produce
it.  A given state can be massless ({\it i.e.}\/, appear in the low-energy
string spectrum) if and only if $h_R\leq 1$.
For example, for $SO(10)$
realized at level $k=2$, the $\rep{54}$ representation has $h=1$.
Since this saturates the string masslessness constraint,
no additional string excitations
are permitted for this representation beyond those that produced
the $\rep{54}$ quantum numbers.
In other words, in any level-two $SO(10)$ string model,
the $\rep{54}$ representation must be a singlet under
all gauge symmetries beyond $SO(10)$.
Similar conclusions also hold for the $\rep{210}$ representation of
$SO(10)$ at level four.
In both cases,
such string constraints have profound
consequences for the allowed couplings or superpotential
terms that involve these representations.

The above sorts of results are easy to derive, and have been
explored at great length in the literature.
For example,
a particularly detailed analysis of the constraints following
from Table~\ref{tableone} can be found in Ref.~\cite{eln}.
However, all such constraints are based purely on considerations of the affine
Lie algebras themselves, such as their central
charges and conformal dimensions.  Indeed, except
for the simple string requirements $c\leq 22$ and $h\leq 1$,
they do not rely on any deeper aspects of the methods by which
such algebras are ultimately {\it realized}\/
within the physical context of string theory.
In this paper, however, we shall show that there are additional
constraints which arise precisely from
such physical considerations in the case of free-field string models.
Such free-field string constructions include those based upon free worldsheet
bosons and fermions (such as the free-fermionic, lattice, or orbifold
constructions \cite{constructions}),
and encompass all of the string constructions which have formed the
basis of string GUT model-building attempts in recent years.
As shown in Ref.~\cite{embeddings},
the methods by these higher-level gauge symmetries
can actually be realized in such string constructions are highly limited.
As a consequence, we shall be able to derive a variety of
new constraints which are so powerful that they effectively
replace the above table
with a new, substantially more restrictive one.

These new methods that we shall use have already proven to be quite powerful
when applied to higher-level $SO(10)$ free-field string GUT models.
For example, it was recently shown \cite{embeddings} that
despite the central charge constraint which na\"\i vely permits
$SO(10)$ to be realized at levels $k\leq 7$,
in actuality $SO(10)$ cannot be realized beyond level {\it four}\/ in such
constructions.
Thus, Table~\ref{tableone} provides in fact a complete list of potentially
allowed representations.
This in turn provides the first proof, for example, that the $\rep{126}$
representation of $SO(10)$ can never be realized in
free-field heterotic string theory \cite{embeddings}.

In this paper, we shall use the methods of Ref.~\cite{embeddings} to obtain
a number of further striking results.
First, just as the
level-two $\rep{54}$ representation transforms as a singlet under all gauge
symmetries
beyond $SO(10)$, we shall show that
in string models employing the so-called ``diagonal embeddings''
(which includes all string GUT models constructed
to date),
in fact {\it all}\/ representations larger than the $\rep{16}$ representation
must
also transform this way.
Indeed, we shall show that this result holds {\it regardless}\/
of the affine level at which the $SO(10)$ gauge group is realized.
In other words, such representations always behave as though
they have {\it effective}\/ conformal dimensions $h_{\rm eff}=1$.
Second, just as the $\rep{126}$ representation cannot be obtained
in free-field string models, we shall show in fact that
all representations higher than the $\rep{54}$
cannot be obtained
via diagonal embeddings --- again, regardless of the level.
This includes the useful $\rep{120}$ and $\rep{144}$ representations,
which one might have expected to arise at levels $k=3$ or $k=4$.
Indeed, the only exception to this rule will be the level-four
diagonal embedding, which is capable of yielding a $\rep{210}$ representation.
We shall also examine the case of $E_6$ string models,
and prove that similar results hold for this case as well:
all adjoint $\rep{78}$ representations
must transform as singlets under all gauge symmetries beyond $E_6$,
and all representations beyond the $\rep{78}$ representation are prohibited.

All of these results are completely model-independent, and do not
depend on the particular subclass of free-field string model construction
that is used ({\it e.g.}, free fermions, orbifolds, lattices, {\it etc}.).
Taken together, then, these
results place severe restrictions on the types of field-theoretic
$SO(10)$ and $E_6$ GUT models that can be derived from
free-field heterotic string theory using diagonal embeddings.
For example, in the case of $SO(10)$,
they essentially rule out the models in Ref.~\cite{bads}, and
favor the models in Ref.~\cite{goods}.
Given these results, we shall then also explore possible string constructions
using {\it non}\/-diagonal embeddings,
and show that in the $SO(10)$ case,
the phenomenological prospects can be significantly improved.

This paper is organized as follows.
In the next section, we shall first review the results
of Refs.~\cite{embeddings,embeddings2} that will be necessary for our
subsequent analysis.
Then, in Sect.~3,
we will apply these techniques to the case
of $SO(10)$ string GUT models built using diagonal embeddings.
The case of non-diagonal $SO(10)$ constructions will
be considered in Sect.~4, and in Sect.~5 we will turn our
attention to $E_6$ string GUT models.  Finally, in Sect.~6, we
will summarize our conclusions and discuss some of
their phenomenological ramifications.

\section{Allowed SO(10) Embeddings and Matter Representations}

Let us first begin by briefly reviewing the techniques
that underlie our analysis.
As we have stated, these techniques are ultimately a consequence
of the methods by which free-field string theories yield
not only higher-level gauge symmetries, but also their corresponding
matter representations.
These methods are discussed in Refs.~\cite{embeddings,embeddings2}
respectively.  We shall now review the salient points for each.

\subsection{Allowed SO(10) embeddings}

In Ref.~\cite{embeddings},
a general examination of the methods
by which higher-level gauge symmetries can be realized in string theory
was conducted.  The main results can be summarized as follows.
First, it was shown in Ref.~\cite{embeddings} that
conformal invariance and
the masslessness constraint for gauge boson states together imply
that the only way in which a higher-level gauge symmetry $G'$ at level $k$ can
be realized in free-field heterotic string theory is to start with
a certain level-one simply laced gauge symmetry (henceforth denoted $G$) with
rank
$\leq 22$, and then to perform a so-called {\it dimensional truncation}\/ of
its charge
lattice via a GSO projection.
Thus, one can realize a higher-level gauge symmetry in free-field string theory
only by first building a string model which realizes the larger level-one gauge
group
$G$, and then truncating to the higher-level subgroup $G'$.
Second, it
was also shown \cite{embeddings} that each such dimensional truncation
corresponds uniquely to a so-called  {\it irregular}\/ group-theoretic
embedding  $G'_k\subset G$ (where `irregular' signifies that some
of the roots of the subgroup are not roots of the original group).
Since these irregular group-theoretic embeddings have been classified
by mathematicians,
these two observations together permitted a
complete classification \cite{embeddings} of the methods by which
all higher-level gauge symmetries can be realized in free-field string theory,
and the results for the case of $SO(10)$ are given
in Table~\ref{tableembeddings}.
We shall defer our discussion of $E_6$ to Sect.~5.

\begin{table}[t]
\centerline{
\begin{tabular}{c|rcl|c|c}
 Label & \multicolumn{3}{c|}{Embedding} &  $\Delta r$ & $\Delta c$\\
\hline
\hline
  I & $SO(10)_2$  & $\subset$ &   $[SO(10)_1]^2$       &  $5$ & $1$ \\
  II & $SO(10)_2$  & $\subset$ &   $SU(10)_1$           &  $4$ & $0$ \\
\hline
  III & $SO(10)_3$  & $\subset$ &  $[SO(10)_1]^3$        &  $10$ & $30/11$ \\
\hline
  IV & $SO(10)_4$  & $\subset$ &  $[SO(10)_1]^4$        &  $15$ & $5$ \\
  V & $SO(10)_4$  & $\subset$ &  $SU(16)_1$            & $10$ & $0$ \\
  VI & $SO(10)_4$  & $\subset$ &  $[SU(10)_1]^2$        & $13$ & $3$  \\
  VII & $SO(10)_4$  & $\subset$ &  $SU(10)_1 \times [SO(10)_1]^2$
                                                   & $14$ & $4$ \\
\hline
  ~ &   $ SO(10)_{k>4} $  & &  Impossible & ~  & ~  \\
\hline
\end{tabular}
 }
\caption{Complete list of string-allowed embeddings for $SO(10)$
     at higher levels, where $\Delta r\equiv {\rm rank}\,G-5$
      and $\Delta c\equiv c(G)-c(SO(10)_k)$.  The embeddings have
     been labelled I through VII for future reference.}
\label{tableembeddings}
\end{table}

Note that the embeddings listed in Table~\ref{tableembeddings}
are the {\it only}\/ ways in which $SO(10)$ at higher levels can
be realized in free-field string constructions.
Thus, we see that when constructing $SO(10)$ at higher levels,
the cost we must pay is not simply that of realizing
the $SO(10)$ group factor itself;
rather, we must start by actually realizing the larger embedding group $G$,
and thereby pay additional costs both in rank $\Delta r$ and
in central charge $\Delta c$.
When $\Delta c\not= 0$,
these extra costs have a profound effect, for they result in the
generation of extra unwanted non-gauge chiral algebras
of central charge $\Delta c$
in addition to the desired higher-level gauge symmetry \cite{embeddings}.

Note that embeddings of the form
\beq
        SO(10)_k ~\subset ~[SO(10)_1]^k
\eeq
are the traditional so-called {\it diagonal}\/ embeddings.
Indeed, most string GUT models constructed to date employ such embeddings:
one first realizes the gauge symmetry $[SO(10)_1]^k$ in a given string model,
and then breaks the gauge symmetry down to the higher-level diagonal subgroup
$SO(10)_k$
(for example, by modding out by the permutation symmetry between the
different $SO(10)$ gauge factors).
By contrast,
one of the main achievements
of Ref.~\cite{embeddings} was to identify all of the other {\it non}\/-diagonal
embeddings which can also produce $SO(10)$ at higher levels.
As evident from Table~\ref{tableembeddings},
these non-diagonal embeddings are always more efficient than the diagonal
embeddings,
with smaller $\Delta r$ and $\Delta c$.
Indeed, those embeddings with $\Delta c=0$ are called ``conformal'', and will
prove to be extremely useful in what follows since they do not generate any
unwanted chiral algebras.
Finally, also note that there do not exist any embeddings yielding $SO(10)_k$
for
$k>4$.  Hence such levels cannot be realized in free-field string
constructions.

\subsection{Allowed SO(10) matter representations}

While the discussion in Ref.~\cite{embeddings} focused
on the realization of the $SO(10)$ gauge boson states, in this paper
we shall use the results in Table~\ref{tableembeddings}
in order to study the possible $SO(10)$ {\it matter}\/ representations.
In general, just as was the case for the gauge boson states,
a given matter representation of
$SO(10)$ at any level is permitted
if and only if it can also be realized through the embeddings in this table.
There are, however, certain subtleties which must be taken into account.
In particular, for each embedding $SO(10)_k\subset G$, a given matter
representation
can arise in either an untwisted or a twisted string sector.
It turns out that the rules for analyzing the states from the
untwisted and twisted sectors may differ depending on the particular
embedding $SO(10)_k\subset G$ under study.
However, as derived and discussed in Ref.~\cite{embeddings2},
there exists a simple rule which governs this process.

In order to discuss this rule,
let us first introduce some terminology.
We shall say that a string model realizes a given
embedding $G'_k\subset G$ if, as above, this is the embedding through
which the {\it gauge bosons}\/ of $G'_k$ are realized.
Furthermore, when we label the various sectors of a given string
model as either ``untwisted'' or ``twisted'', these adjectives will
refer to the twist that induced the final breaking
$G\supset G'_k$.
Finally, we shall
refer to a given matter representation $R'$ of any group $G'_k$
as being ``allowed by the embedding'' $G'_k\subset G$ if
and only if it arises
as the result of the decomposition of a unitary potentially massless
representation $R$ of $G$.  In other words, there should exist a branching
rule of the form $R\to R'\oplus ....$ where $R$ is a unitary representation
of $G$ with conformal dimension $h\leq 1$.

Given this terminology,
the rule for describing the allowed
potentially massless matter representations of $SO(10)_k$
can be easily summarized as follows \cite{embeddings2}.
If a given model realizes
an embedding $SO(10)_k\subset G$,
then it turns out that
a given matter representation of $SO(10)_k$
can arise
in the corresponding {\it untwisted}\/ string sectors
only if it is allowed by the same embedding $SO(10)_k\subset G$.
Moreover, if this embedding $SO(10)_k\subset G$ is
non-conformal (so that $\Delta c\not= 0$),
then this restriction must apply
to the states from the {\it twisted}\/ string sectors as well --- {\it i.e.}\/,
they too must be allowed by the same embedding $SO(10)_k\subset G$.
Thus, for string models that realize non-conformal embeddings,
the states from the twisted and untwisted sectors behave similarly,
and must arise from decompositions of unitary potentially massless
representations of the same original group $G$.

By contrast, if the original gauge-group embedding $SO(10)_k\subset G$
is {\it conformal}\/,
then it turns out that
there is no model-independent constraint that correlates the effective
embeddings describing the origins of the twisted-sector
matter states with those describing the origins of the untwisted-sector matter
states.
In other words, while the matter states in the untwisted sectors
must still be those that are allowed by the conformal embedding
$SO(10)_k\subset G$,
the matter states from the twisted sectors
can in general be those that are allowed by {\it any other}\/ embedding
$SO(10)_k\subset H$
in Table~\ref{tableembeddings} with the same value of $k$.
Thus, in general, these twisted-sector states can behave as though they
were the {\it untwisted}\/ states relative to a different (not necessarily
conformal)
effective embedding $SO(10)_k\subset H$.
Note that we are not claiming that such effective embedding
``mixtures'' will actually arise in a given model --- we are merely claiming
that
they are are {\it allowed in principle}.  Whether they actually arise depends
on the details of the particular string construction.

It is easy to see why the conformal and non-conformal cases
are so different.
As explained in Ref.~\cite{embeddings2},
the crucial point to consider is the tree-level coupling
of the $SO(10)$ gauge bosons to the matter representations
that carry $SO(10)$ charges.  Such couplings are characterized
by the three-point matter/matter/gauge-boson vertices in the
low-energy effective field theory.
For consistency, such couplings must be non-vanishing regardless
of whether these matter representations arise in twisted or
untwisted strings sectors.
However, this has different implications depending on whether
the gauge boson embedding is conformal or non-conformal.

If the gauge boson embedding is non-conformal,
then an extra chiral algebra will be generated in addition
to the higher-level gauge symmetry.
This implies that the full vertex operator for the gauge
boson states will contain not only the usual gauge contribution, but also
a contribution from the extra chiral algebra.
This in turn implies that the gauge bosons cannot couple to charged matter
representations
unless the vertex operators
of such matter states also contain contributions from the same chiral algebra
(so that the two contributions may cancel to permit a non-vanishing
correlation function).
Of course, by definition, this will be the case for matter
arising in the untwisted string sectors.
However, for matter arising in the twisted string sectors,
such couplings cannot exist unless the vertex operators of these states
also contain contributions from the same chiral algebra.
In other words, it is impossible to couple
the gauge bosons to charged twisted-sector matter states unless
these twisted-sector states also behave as though they are realized via
the same embedding.
This is ultimately the origin of the rule discussed above.

By contrast,
for gauge bosons realized via conformal embeddings, the situation is more
flexible
because in such cases the vertex operators of the gauge boson states do not
carry any
contributions from extra chiral algebras.  Thus, they can generally couple to
charged matter representations from all sectors regardless of
the particular effective embedding that may be realized in the twisted sectors.
Note that this is true even if the effective embedding describing the
twisted-sector
states is non-conformal.  Indeed, in such cases,
the twisted-sector matter representations carry their own
chiral-algebra contributions,
but these nevertheless cancel between themselves
in the three-point matter/matter/gauge-boson vertices.
These rules and their ramifications are discussed more fully in
Ref.~\cite{embeddings2}.

Thus, for each higher-level $SO(10)$ embedding listed in
Table~\ref{tableembeddings},
we can now determine the properties of the corresponding potentially massless
$SO(10)_k$ matter representations simply by determining those that
are ``allowed'' --- {\it i.e.}, by determining those which are realizable via
the
decompositions of potentially massless representations of
the appropriate original groups $G$ according the above rules.
This is the procedure we shall follow in this paper.

\section{SO(10):  Analysis for Diagonal Embeddings}

To illustrate, let us start by considering the first embedding in
Table~\ref{tableembeddings},
namely the level-two diagonal embedding $SO(10)_2\subset [SO(10)_1]^2$.
We see from Table~\ref{tableone} that the only allowed potentially massless
representations of $SO(10)_1$ are the $\rep{10}$ and $\rep{16}$
representations,
with conformal dimensions $h=1/2$ and $5/8$ respectively.
Thus, the only allowed distinct non-singlet potentially massless
representations of $[SO(10)_1]^2$ are the $(\rep{1},\rep{10})$,
$(\rep{1},\rep{16})$, and $(\rep{10},\rep{10})$ representations,
with conformal dimensions $h=1/2$, $5/8$, and $1$ respectively.
Under the diagonal embedding, these representations
respectively decompose into
the $\rep{10}$, $\rep{16}$, and $\rep{1}\oplus \rep{45}\oplus \rep{54}$
representations of $SO(10)_2$.
Thus, in order to realize the $\rep{45}$ or $\rep{54}$ representations
of $SO(10)_2$ via this embedding, we see that we must first realize the
$(\rep{10},\rep{10})$ representation of $SO(10)_1\times SO(10)_1$,
and then do the appropriate decomposition.

Note, however, that each of these final representations inherits
the conformal dimension of its parent representation.
Indeed, the only difference is that after the decomposition,
 {\it only part}\/  of this conformal dimension
can be attributed to the $SO(10)_2$ gauge affine Lie algebra itself;
the remaining contribution arises from the additional
chiral algebra.
For example, since the original $(\rep{10},\rep{10})$
representation has $h=1$, the final $\rep{45}$ representation
that it produces also inherits this conformal dimension:
one contribution $h=4/5$ arises from the
$SO(10)_2$ gauge affine algebra itself, and the remaining contribution
$h=1/5$ arises from the non-gauge chiral algebra.
Indeed, $h=1/5$ is the extra ``cost'' of realizing
the $\rep{45}$ representation in this construction.
A similar observation for this case appears in Ref.~\cite{AFIU}.
In fact, since the only way to realize the $\rep{45}$
representation in this embedding is via
the $(\rep{10},\rep{10})$ representation, the $\rep{45}$
representation is correlated with a unique
set of excitations of the primary fields of the chiral
algebra.
This is in fact true of all of the $SO(10)_2$ representations
in this embedding, since each
representation has a unique origin in the pre-truncated theory.
Thus, for these reasons, we may consider each of the original
conformal dimensions to be a true ``effective'' conformal dimension,
so that the
$\rep{10}$, $\rep{16}$, $\rep{45}$, and $\rep{54}$ representations
of $SO(10)_2$ in this embedding
have effective conformal dimensions $h_{\rm eff}=1/2$, $5/8$, $1$, and
$1$ respectively.

Note that this is a major
change beyond what appears in Table~\ref{tableone}.
For example, even though the adjoint $\rep{45}$ representation at
level two has a true conformal dimension $h=4/5$,
we now see that it nevertheless has an effective conformal dimension
$h_{\rm eff}=1$ in this embedding.
Thus, there cannot be any additional excitations that could provide this
representation with the quantum numbers of any other gauge groups!
Indeed, in this respect, it resembles the $\rep{54}$ representation which is
also a singlet (in this case under the chiral algebra as well as
under other gauge groups).
Thus, we see that any $\rep{45}$ or $\rep{54}$ representations
produced via this embedding will be singlets under all other gauge groups.

This result holds for the level-three and level-four diagonal embeddings
as well.
In the case of the level-three diagonal embedding $SO(10)_3\subset
[SO(10)_1]^3$,
the only allowed potentially massless non-singlet representations of
$[SO(10)_1]^3$
are the $(\rep{1},\rep{1},\rep{10})$, $(\rep{1},\rep{1},\rep{16})$,
and $(\rep{1},\rep{10},\rep{10})$ representations, along with their
permutations.
These have conformal dimensions $h=1/2$, $5/8$, and $1$ respectively,
and have branching rules similar to those of the level-two embedding.
Thus they too can only give rise to
the $\rep{10}$, $\rep{16}$, $\rep{45}$, and $\rep{54}$ representations
of $SO(10)_3$,
with effective conformal dimensions $h_{\rm eff}=1/2$, $5/8$, $1$, and
$1$ respectively.
The same result also holds for the level-four embedding.

Does this imply that these four representations are
the {\it only}\/ matter representations that can ever be realized in
higher-level
$SO(10)$ string models using diagonal embeddings?
It may seem that there are two objections.
First, as discussed in Sect.~2, it may seem that the above line
of reasoning should only apply to the {\it untwisted}\/ string sectors.
After all, while we are considering the effects
of the dimensional truncation on the sectors that were present prior to the
twist that induces the dimensional truncation,
it may seem that we are ignoring the entirely {\it new twisted}\/ sectors that
modular invariance forces us to add to the string model.
However,
any states which arise in the twisted sectors must also
obey the same effective diagonal decomposition rules
as do the untwisted-sector states.
As we have discussed in Sect.~2, this correlation arises because
the diagonal embeddings are non-conformal, with $\Delta c \not= 0$.
Indeed, as shown in Ref.~\cite{embeddings2},
if the twisted-sector states were to arise in some other manner
({\it e.g.}\/, via a different effective embedding $SO(10)_k\subset G$),
then the gauge bosons from the $SO(10)_k\subset [SO(10)_1]^k$ embedding
could not couple to the charged $SO(10)$ representations from the twisted
sectors, for the gauge-boson vertex operators would contain contributions
from the $[SO(10)_1]^k/SO(10)_k$
chiral algebra that would not be cancelled by similar
contributions from the vertex operators of the twisted-sector states.
This failure of the $SO(10)$ gauge bosons to couple to the charged
twisted-sector matter
states would then signal an inconsistency in the underlying string
construction.
Thus, the above analysis must hold for both the untwisted and twisted sectors.

The second possible objection to the above results concerns the fact
that when studying the decompositions of the allowed representations
of $[SO(10)_1]^k$, we used the branching rules
of the Lie algebra decomposition $SO(10)^k\supset SO(10)$
rather than those of the {\it affine}\/ Lie algebra decomposition
$[SO(10)_1]^k\supset SO(10)_k$.
This is an important distinction,
for while the ordinary Lie algebra branching rules are
sufficient for understanding the decomposition properties
of the {\it finite}\/-dimensional group representations,
in our case we actually must deal with the
 {\it infinite}\/-dimensional affine Lie algebra representations
which also include the infinite Virasoro towers of states.
Indeed, while the use of the finite branching rules
is sufficient for those $SO(10)_k$ primary fields that arise from the primary
fields
of $[SO(10)_1]^k$ (and which correspond to the lowest-lying representations
in each Virasoro tower of states),  it is possible
that $SO(10)_k$ primary fields
can also arise from the {\it descendent}\/ fields of $[SO(10)_1]^k$ after
the $[SO(10)_1]^k \supset SO(10)_k$ decomposition.
This would correspond to the $[SO(10)_1]^k$ towers of states
reshuffling and recombining to fill out
$SO(10)_k$ towers of states
in such a way that
a descendent (excited) state in the original Verma module
of the original algebra
becomes a primary (lowest-lying) state in the final Verma module.
As discussed in Ref.~\cite{embeddings2},
such extra possibilities would be recorded within the full {\it affine}\/
branching rules.
Thus, in order to investigate this possibility, one must first derive
the affine branching rules for the diagonal embeddings $SO(10)_k\subset
[SO(10)_1]^k$.

In general, such affine branching rules are rather difficult to derive,
and only partial results exist in the mathematical literature for general
embeddings $G'\subset G$ \cite{fuchs}.
Indeed, these branching rules might seem to be especially complicated
in our case since the values of $\Delta c$ that we face are larger than one,
and hence the possible chiral conformal field theories
whose representations must be tensored with the Verma modules of $G'$
have not even been classified.\footnote{
       Note that for the level-two diagonal $SO(10)$ embedding,
       we in fact have $\Delta c=1$.  In this case
       the extra chiral algebra corresponds to that of a boson
       on a ${\bf Z}_2$ orbifold of radius $\sqrt{10}/2$.}
Fortunately, however,
there are two observations that make the job simpler.

First, it is possible to use $SO(10)$ charge conservation
in order to quickly constrain the allowed representations that could possibly
arise via such descendent fields of $[SO(10)_1]^k$.
Note that the only potentially massless
representations of $[SO(10)_1]^k$ that concern us
are the $(\rep{1,1})$, $(\rep{1},\rep{10})$, $(\rep{1},\rep{16})$,
and $(\rep{10},\rep{10})$ representations (with a sufficient number of
extra singlet representations attached if $k>2$).
As discussed in Ref.~\cite{embeddings2}, $SO(10)$ charge conservation
then implies that the only descendent representations of $SO(10)_k$ that
could possibly arise from these representations are as follows:
$\rep{45}$, $\rep{54}$, and $\rep{210}$  representations from
$(\rep{1,1})$ and
$(\rep{10},\rep{10})$;
a $\rep{120}$ representation from $(\rep{1},\rep{10})$;
and a $\rep{144}$ representation from $(\rep{1},\rep{16})$.
Of course, after the $[SO(10)_1]^k\supset SO(10)_k$ decomposition,
these representations can only appear as primary fields
if they are permitted according to Table~\ref{tableone}.

Second, this situation simplifies even further if we now
recall that
descendent fields have conformal dimensions which always differ
by positive integers from those of their underlying primary fields.
Thus, we see that the only possible descendent
fields that can ever arise in the massless string spectrum
must in fact be descendents of the {\it identity}\/
representation of $[SO(10)_1]^k$, since this is the only representation with
$h=0$.

Thus, we reach two very important conclusions.
First, we see that the useful $\rep{120}$ and $\rep{144}$ representations
of $SO(10)$ never arise as ``primary-from-descendent'' fields.
However, since we have already shown that they also cannot arise as
``primary-from-primary'' fields, we learn that they cannot be realized at all!
Thus, these representations of $SO(10)_k$ are essentially ruled out
in diagonal constructions.
Note that this result applies regardless of the affine level
at which $SO(10)$ is realized.
Indeed, we see that
the only ``large'' representation that can ever be realized in
such constructions
is the $\rep{210}$ representation, which can appear only
as a ``primary-from-descendent'' field at level $k=4$.

Second, we see that although
the $\rep{45}$ and $\rep{54}$ representations
of $SO(10)_k$ can now also be
realized as ``primary-from-descendent'' fields,
they nevertheless still must transform as singlets
under all gauge symmetries beyond $SO(10)$.
This is because they can only arise as the first descendents of the identity
field, and will therefore continue to have $h_{\rm eff}=1$.
Indeed, this is a general property of {\it all}\/
potentially massless ``primary-from-descendent''
fields, and applies for both diagonal and non-diagonal embeddings.
Of course, for conformal embeddings, such fields will not only have $h_{\rm
eff}=1$,
but will in fact have $h=1$ (which
renders the identification of such candidate representations quite simple).

Thus, to summarize, we have shown the following.
For $SO(10)$ diagonal embeddings, we see that
the only possible ``primary-from-primary'' fields that can ever
arise are the $\rep{10}$, $\rep{16}$, $\rep{45}$, and $\rep{54}$
representations,
with effective conformal dimensions $h_{\rm eff}=1/2$, $5/8$, $1$, and
$1$ respectively.  This result holds regardless of the affine level
at which $SO(10)$ is realized, and regardless of the particular
free-field string construction employed (whether free-fermions, orbifolds,
lattice compactifications, {\it etc.}\/).
Likewise, the only possible
``primary-from-descendent'' fields that can ever
arise at the massless level are the
$\rep{45}$, $\rep{54}$, and $\rep{210}$
representations, but
once again we have shown that
such representations still must have $h_{\rm eff}=1$,
and must therefore also transform as singlets under all gauge symmetries
beyond $SO(10)$.
Thus, as a general statement, we see that {\it any}\/ representation larger
than the $\rep{16}$ must transform as a singlet under all gauge symmetries
beyond $SO(10)$.
Moreover, the $\rep{120}$ and $\rep{144}$ representations are excluded
in such constructions, regardless of the affine level at which $SO(10)$ is
realized.

Note that while these results hold
only for the diagonal embeddings,
we see from Table~\ref{tableembeddings}
that the diagonal embedding is in fact the {\it only}\/ possible
embedding in the particular case of
$SO(10)$ at level three.
Thus, for $SO(10)_3$, the above results are in fact completely general.
This observation is particularly important, for
the only known three-generation $SO(10)$ string GUT models \cite{KT}
at present are all at level three.
As expected, the spectra of these models agree precisely with our
general conclusions.

\section{SO(10):  Analysis for Non-Diagonal Embeddings}

Given the above results for the diagonal embeddings,
let us now turn to the possible {\it non}\/-diagonal embeddings.
We shall find that in these cases there is a far richer structure,
with many more phenomenologically interesting possibilities.

First, we analyze the non-diagonal level-two embedding,
namely $SO(10)_2\subset SU(10)_1$.
In this case, the only allowed potentially massless unitary
$SU(10)_1$ representations
are the $\rep{10}$ representation with $h=9/20$, and the $\rep{45}$
representation with $h=4/5$.  These  satisfy the $SU(10)_1\supset SO(10)_2$
branching rules $\rep{10}\to\rep{10}$ and $\rep{45}\to\rep{45}$.
Thus, this embedding is capable of producing a $\rep{10}$ representation
with $h_{\rm eff}=9/20$, and a $\rep{45}$ representation with $h_{\rm
eff}=4/5$.
Unlike the diagonal embeddings, this embedding is therefore capable
of producing
a $\rep{45}$ representation which need not transform as a singlet
under all gauge symmetries beyond $SO(10)$.

This embedding, however, can never yield a $\rep{16}$
representation of $SO(10)_2$.
It may seem that this embedding is therefore useless on phenomenological
grounds.  Remarkably, however, it turns out that this embedding is conformal,
with $\Delta c=0$.  This means that it generates no chiral
algebra, which in turn means that there is no constraint that forces
the twisted-sector states to be realized from the same effective embedding
as the untwisted-sector states.
In particular, it is possible that the
twisted-sector states can arise from an effective
 {\it diagonal}\/ embedding, as described above.  Thus, it may be possible to
build
$SO(10)_2$ string models in which the $\rep{16}$ representations
with $h_{\rm eff}=5/8$ come from the twisted sectors, while
the phenomenologically interesting $\rep{45}$ representations come from
the untwisted sectors.
Note that this scenario is possible, however, only if the higher-level gauge
bosons are realized through this non-diagonal embedding.
Also note that the only ``primary-from-descendent'' representation which can
possibly
arise in this case is the $\rep{54}$ representation, since it already has
$h=1$.
This is because no chiral algebra is present which could supply
any missing conformal dimension in order to build $h_{\rm eff}=1$
for any of the other representations.

Finally, we turn to the non-diagonal embeddings at level four.
Considering first the $SO(10)_4\subset SU(16)_1$ embedding, we see that
that the only potentially massless unitary
$SU(16)_1$ representations are the $\rep{16}$ representation with $h=15/32$,
and the $\rep{120}$ representation with $h=7/8$.  These satisfy the simple
branching rules $\rep{16}\to\rep{16}$ and $\rep{120}\to \rep{120}$, from which
we obtain the $\rep{16}$ representation of $SO(10)_4$ with $h_{\rm eff}=15/32$,
and the $\rep{120}$ representation of $SO(10)_4$ with $h_{\rm eff}=7/8$.
Even though no $SO(10)_4$ adjoint or $\rep{54}$ scalar representation is
produced
in such untwisted sectors,
we need not disregard this embedding because it too is conformal, and thus
the states from the twisted sectors may arise as the result of
a different effective level-four embedding
(such as the level-four diagonal embedding).
Also note that, like the above level-two conformal embedding,
the only ``primary-from-descendent'' state that can possibly arise
in this case is the $\rep{210}$ state, since it already
has $h=1$ at level four.

Next, we consider the $SO(10)_4\subset [SU(10)_1]^2$
embedding.
It is straightforward to see, by the same sort of analysis,
that this embedding
produces $\rep{10}$, $\rep{45}$, and $\rep{54}$ representations,
but no $\rep{16}$ representation.
However, since this embedding is non-conformal, the only
possible states from the twisted sectors are the same as those
from the untwisted sectors.
Thus, unless the necessary $\rep{16}$ representations
can somehow arise as ``primary-from-descendent'' fields
(which is unclear given the fact that the required affine branching
rules have not yet been derived in the mathematical literature),
we must reject this embedding on phenomenological grounds.

Finally, we consider the $SO(10)_4\subset SU(10)_1\times [SO(10)_1]^2$
embedding.
In this case,
the only allowed distinct non-singlet potentially massless
representations of $ SU(10)_1\times SO(10)_1\times SO(10)_1$
are as follows:
\beqn
& (\rep{1}, \rep{1}, \rep{10}) &~{\rm with }~ h=1/2\nonumber\\
& (\rep{1}, \rep{1}, \rep{16}) &~{\rm with }~ h=5/8\nonumber\\
& (\rep{1}, \rep{10}, \rep{10}) &~{\rm with }~ h=1\nonumber\\
& (\rep{10}, \rep{1}, \rep{1}) &~{\rm with }~ h=9/20\nonumber\\
& (\rep{45}, \rep{1}, \rep{1}) &~{\rm with }~ h=4/5\nonumber\\
& (\rep{10}, \rep{10}, \rep{1}) &~{\rm with }~ h=19/20~
\eeqn
(along with their permutations).
These respectively decompose into the
$\rep{10}$;
$\rep{16}$;
$\lbrace \rep{45}\oplus \rep{54}\oplus \rep{1}\rbrace$;
$\rep{10}$;
$\rep{45}$;
and
$\lbrace \rep{45}\oplus \rep{54}\oplus \rep{1}\rbrace$
representations of $SO(10)_4$.
Thus, we see that this embedding
succeeds in providing both of the required $\rep{16}$ and $\rep{45}$
representations.  Moreover, we see that the $\rep{45}$ representation
arises  with a range of possible effective conformal dimensions
depending on how it is produced.

\section{Analysis for $E_6$}

Although the above analysis concentrated
on the case where the GUT group is $SO(10)$, the methods used
are completely general and
can thus be applied to any group.
In this regard, the case of $E_6$
is particularly straightforward, for it has been
shown \cite{embeddings}
that the only
possible free-field embeddings yielding $E_6$ at higher levels
are the diagonal embeddings $(E_6)_k\subset [(E_6)_1]^k$ for levels $k=2,3$.
Indeed,
just as it was found that $SO(10)$ cannot be realized
beyond level four in free-field constructions \cite{embeddings}, it turns out
that $E_6$ cannot be realized beyond level three.
This means, {\it a priori}\/, that the only allowed representations
of $E_6$ are
the $\rep{27}$ and $\rep{78}$ representations, with
respective
conformal dimensions $\lbrace 13/21,6/7\rbrace$ for $k=2$,
and $\lbrace 26/45,4/5\rbrace$ for $k=3$.
Thus, all representations larger than the $\rep{78}$ representation
of $E_6$ are prohibited.

Given the fact that the diagonal embeddings are the only
possible embeddings for $E_6$, an analysis along the lines
presented above is particularly simple.
Indeed, we find that
among the ``primary-from-primary'' fields, only the
 $\rep{27}$ representation is possible.
Such a state arises from the $(\rep{1},\rep{27})$ or
$(\rep{1},\rep{1},\rep{27})$ representations of
$[(E_6)_1]^k$, and always has $h_{\rm eff}=2/3$.
As in the case of $SO(10)$,
this result holds regardless of the level.

At first sight,
it may therefore appear that $E_6$ string GUT models
are ruled out because of the absence of the required
adjoint $\rep{78}$ Higgs
representation.
However, this is not the case, for such a representation
can (and in fact does) appear
as a ``primary-from-descendent'' field.
To see that such a representation does appear, let us consider
the level-two diagonal embedding,
$(E_6)_2\subset (E_6)_1\times (E_6)_1$.
This embedding has $\Delta c=6/7$, and therefore
generates an extra chiral algebra which
corresponds to the $m=6$ member of the $c<1$ unitary
minimal-model series.
This conformal field theory, which is associated with the tricritical
three-state Potts model,
has primary fields $\phi_{p,q}$
with conformal dimensions $h_{p,q}= [(7p-6q)^2-1]/168$,
where $1\leq p \leq 5$ and $1\leq q\leq p$.
Note that the identity field
in this conformal field theory is $\phi_{1,1}$.
Under the decomposition
\beq
 (E_6)_1\times (E_6)_1 ~\supset ~
   (E_6)_2 \times \lbrace \, m{=}3 \rm \,~minimal\,~model\,\rbrace ~,
\eeq
one can then show that
the full affine branching rule for the identity representation
$(\rep{1},\rep{1})$ of $(E_6)_1\times (E_6)_1$ is
\beq
  (\rep{1},\rep{1}) ~\to~
  (\rep{1},\phi_{1,1}) ~\oplus~ (\rep{78},\phi_{5,5})~.
\eeq
Note that the $\rep{78}$ representation of $(E_6)_2$
has $h=6/7$,
so that the
full $(\rep{78},\phi_{5,5})$ representation indeed has $h=1$.
Thus, as expected, we see that the $\rep{78}$ primary field of $(E_6)_2$ arises
as the first descendent of the identity field of
$(E_6)_1\times (E_6)_1$, and has $h_{\rm eff}=1$.
Moreover, we see from this analysis that this is the
 {\it only}\/  way of realizing an adjoint $\rep{78}$ scalar
in $E_6$ string GUT models.
Conversely, we also see that the $\rep{27}$
representation of $(E_6)_2$ {\it cannot}\/  arise this way.

Thus, in the case of $E_6$,
we have proven that all $\rep{27}$ representations must
in fact have $h_{\rm eff}=2/3$;  that all $\rep{78}$ representations
must in fact have $h_{\rm eff}=1$;
and that all larger representations are prohibited.
This implies, as in the case of $SO(10)$, that
despite our na\"\i ve
expectations based on considerations of the
affine Lie algebra alone (which would have seemed to permit
extra gauge quantum numbers for both the $\rep{27}$
and $\rep{78}$ representations),
the largest possible representation
is in fact the adjoint $\rep{78}$ representation,
and it must always transform
as a singlet under all gauge symmetries beyond $E_6$.
This has indeed been found to be the case
in all $E_6$ string models constructed
to date \cite{erler,KT}, and seriously restricts the allowed
low-energy phenomenologies
that such $E_6$ string GUT models can have.

Note that similar analyses for $SU(5)$, $SU(6)$,
and other GUT groups of interest can also be undertaken
along the lines presented here.
The results for such cases will be presented elsewhere \cite{embeddings2}.

\section{Conclusions and Discussion}

The results of our analysis for the case of $SO(10)$
are collected in Table~\ref{tabletwo},
where we have listed the only representations of $SO(10)$ that can
actually appear in free-field heterotic string models, along with
their ``effective'' conformal dimensions.
Thus, this new Table~\ref{tabletwo} --- rather than the `standard'
Table~\ref{tableone} --- is the true reflection of the
allowed possibilities for $SO(10)$ string GUT
model-building using free-field constructions.
The analogous modifications to the
`standard' $E_6$ results
were discussed in the previous section.

Perhaps the most important difference between
Table~\ref{tableone} and Table~\ref{tabletwo}
concerns
the increase in the effective
conformal dimensions in the latter
relative to the conformal dimensions in the former.
As we have discussed, this is a reflection of the true ``cost''
of realizing higher-level gauge symmetries
in free-field string models.
In particular, note that the $\rep{45}$ and $\rep{54}$ representations
of $SO(10)$
have $h_{\rm eff}=1$ for each of the diagonal embeddings.
Thus, as we have explained,
such $\rep{45}$ and $\rep{54}$ representations
must always be singlets under all gauge symmetries beyond $SO(10)$.
Also note that despite the appearance of
representations larger than $\rep{54}$ in Table~\ref{tableone},
we see from Table~\ref{tabletwo} that such representations are
indeed quite difficult to obtain.
In fact,
other than the $\rep{210}$ representation which appears
as a ``primary-from-descendent'' field at level four,
we see that no $SO(10)$ representations
larger than the $\rep{54}$ representation can ever appear for diagonal
embeddings,
regardless of the affine level.
Furthermore, even among the non-diagonal embeddings, such representations
appear in only one case:  the conformal embedding $SO(10)_4\subset SU(16)_1$.
This is because the ``cost'' for realizing these representations
is pushed too high in all other cases, and consequently they appear only at the
Planck
scale.

\begin{table}[t]
\centerline{
\begin{tabular}{l|l||l||l|l|l|l}
  \multicolumn{2}{c||}{$k=2$} & ~~$k=3$ & \multicolumn{4}{c}{$k=4$} \\
\hline
  ~~~~~I & ~~~~~~II & ~~~~III & ~~~~~IV & ~~~~~~V  & ~~~~VI & ~~~~VII  \\
\hline
\hline
           (\rep{10},~1/2) &
           (\rep{10},~9/20) &
              (\rep{10},~1/2) &
                       (\rep{10},~1/2) &
                       (\rep{16},~15/32) &
                       (\rep{10},~9/20) &
                       (\rep{10},~9/20)  \\
           (\rep{16},~5/8) &
           (\rep{45},~4/5) &
              (\rep{16},~5/8) &
                       (\rep{16},~5/8) &
                       (\rep{120},~7/8) &
                       (\rep{45},~4/5) &
                       (\rep{10},~1/2) \\
           (\rep{45},~1) &
           ~ &
              (\rep{45},~1) &
                       (\rep{45},~1) &
                       ~ &
                       (\rep{45},~9/10) &
                       (\rep{16},~5/8)  \\
           (\rep{54},~1) &
           ~ &
              (\rep{54},~1) &
                       (\rep{54},~1) &
                       ~ &
                       (\rep{54},~9/10) &
                       (\rep{45},~4/5) \\
           ~ &
           ~ &
              ~ &
                       ~ &
                       ~ &
                       ~ &
                       (\rep{45},~19/20) \\
           ~ &
           ~ &
              ~ &
                       ~ &
                       ~ &
                       ~ &
                       (\rep{45},~1) \\
           ~ &
           ~ &
              ~ &
                       ~ &
                       ~ &
                       ~ &
                       (\rep{54},~19/20) \\
           ~ &
           ~ &
              ~ &
                       ~ &
                       ~ &
                       ~ &
                       (\rep{54},~1) \\
\hline
\end{tabular}
  }
\caption{
The ``new'' complete list of unitary, potentially massless representations of
$SO(10)$
which can ever actually appear in free-field heterotic string models.
Each representation $R$ is listed as $({\bf dim\,R},h_{\rm eff}^{(R)})$
where $h_{\rm eff}^{(R)}$ is its {\it effective}\/ conformal dimension.
In this table, the embeddings are labelled
I through VII
as in Table~\protect\ref{tableembeddings}.
Note that this table does not include possible ``primary-from-descendent''
fields, which (if present) must always have $h_{\rm eff}=1$.
For Embeddings I and III,
no additional ``primary-from-descendent'' fields arise;
likewise, the only possible ``primary-from-descendent'' field for Embedding IV
is
the $\protect\rep{210}$ representation.
For all embeddings except II and V,
the allowed states from the twisted and untwisted sectors are the same.
In the case of Embeddings II and V,
only the states allowed in untwisted sectors are shown;
the states allowed from twisted sectors
can then be any of those listed under Embeddings $\lbrace {\rm I,II}\rbrace$
or $\lbrace {\rm IV,V,VI,VII}\rbrace$ respectively.
   }
\label{tabletwo}
\end{table}

The fact that the diagonal embeddings force the $\rep{45}$
and $\rep{54}$ representations
to be singlets all gauge symmetries beyond $SO(10)$
has a number of undesirable phenomenological consequences.
First, such singlet $\rep{45}$ representations make the doublet-triplet
splitting mechanism hard to implement.
Second, starting from any desired coupling,
they also tend to generate many new
unwanted couplings which arise from
unconstrained insertions/deletions
of $\rep{45}$ and $\rep{54}$ singlets.
Finally, their singlet nature renders it difficult to
construct non-vanishing antisymmetric combinations
of multiple $\rep{45}$ representations,
as are often required \cite{effective126}
in order to simulate the effect of the $\rep{126}$ representation usually
used for the Georgi-Jarlskog mass relations \cite{GJ}.
Indeed, the basic idea here is that an effective
$\rep{126}$ representation can be realized in the higher-order superpotential
as the totally antisymmetric component of a $\rep{45}\cdot \rep{45'}\cdot
\rep{10}$
tensor product, but this requires the presence of two $\rep{45}$
representations
which must differ in some of their additional quantum numbers in order to
permit antisymmetrization.

How then might such antisymmetric couplings be obtained within the diagonal
embeddings?  From the above analysis, we
now see that such antisymmetric couplings can arise
only if two or more $\rep{45}$ representations can be generated
which differ either in their {\it right-moving}\/
string quantum numbers, or in their quantum numbers under the
left-moving chiral algebra.
Unfortunately,
the first option is generally difficult to arrange (since it turns out
to be correlated with the number of generations in the string model),
and the second option is not available at all at level $k=2$
[since all $\rep{45}$ representations at level two originate
from the same parent $(\rep{10},\rep{10})$ representation of $SO(10)^2$].
Indeed, this second option is possible only within the $k=3,4$ diagonal
embeddings
because of the possibility of  permuting the
factors in the corresponding parent representations
$(\rep{1},\rep{10},\rep{10})$ and $(\rep{1},\rep{1},\rep{10},\rep{10})$.
In such cases, however, these permutations only respectively yield
additional ${\bf Z}_3$ and ${\bf Z}_6$
discrete symmetries for the adjoint scalars.

In some sense, these stringent constraints arise for the diagonal
embeddings precisely because they are so
uneconomical.  Indeed, we see from Table~\ref{tableembeddings}
that these embeddings are the most expensive, and require
the largest possible values of $\Delta r$ and $\Delta c$.
It is this which gives rise to the large extra
chiral algebras which lead to these strong constraints.
By contrast, the non-diagonal embeddings are generally far more
efficient, and can yield a greater variety of massless representations
with smaller associated costs.

Thus, as model-builders, it seems that we are faced with two options.
The first is to continue to use the diagonal embeddings,
and to attempt to construct realistic low-energy
field-theoretic $SO(10)$ GUT scenarios
which can be accommodated within the above constraints.
Indeed, the only such field-theoretic $SO(10)$ models which
survive are those, such as in Ref.~\cite{goods}, which
do not make use of the $\rep{120}$, $\rep{126}$,
or $\rep{144}$
representations and which
employ few (preferably only one unique) $\rep{45}$ or $\rep{54}$
representation.
As pointed out
in Ref.~\cite{AFIU}, however, the scenarios in Ref.~\cite{goods}
may run into serious problems because they require explicit mass terms
in the superpotential,
and such mass terms cannot arise in general $SO(10)$ string models \cite{AFIU}.
The second option, by contrast,
is to use the non-diagonal embeddings we have discussed here,
for we see that
they are not only more economical but
also have a far richer structure.
In particular, we see that such embeddings appear to have the
flexibility to give rise to non-singlet adjoint scalars, as well as
possible higher-dimensional representations.
Until now,
these non-diagonal string embeddings
have been little explored,
but our results for the case of $SO(10)$
demonstrate that it is perhaps via these embeddings
that phenomenologically interesting string GUT models may be found.

  \bigskip
  \medskip
  \leftline{\large\bf Acknowledgments}
  \medskip
  I would like to thank S. Barr, G. Cleaver, J. Erler,
   A. Hanany, C. Kolda, J. Lykken,
  F. Wilczek, and E. Witten for discussions.
  I am particularly grateful to
    K.S. Babu, Z. Kakushadze, E. Kiritsis, and S.-H.H. Tye for
   many insightful comments throughout the course of this work,
   and to J. March-Russell for collaborations
(Refs.~\cite{embeddings,embeddings2})
   which supplied many of the general arguments which have been exploited for
the
   $SO(10)$ and $E_6$ analyses presented in this paper.
    I also wish to thank the CERN Theory Group for hospitality while portions
    of this paper were written.
  This work was supported in part by DOE Grant No.\ DE-FG-0290ER40542.

\vfill\eject
  \leftline{\large\bf Note Added}
  \medskip
   After this paper was completed, it was pointed out \cite{zurab} that
   there is slight loophole in one of the arguments that was used
   for restricting the matter content arising from the twisted sectors.
   In certain string models with special patterns of gauge symmetry
   breaking, the states from twisted sectors cannot necessarily be
   described as the untwisted states relative to different effective
   level-one embeddings, and more complicated embedding patterns for
   such states may emerge.  This issue will be discussed fully in
   Ref.~\cite{embeddings2}.  While no known higher-level
   chiral string GUT models exhibit this phenomenon (and therefore all
   known chiral models satisfy the exact constraints we have formulated in
   this paper), there apparently exists a {\it non-chiral}\/ $SO(10)_3$
   model which makes use of this loophole to yield adjoint
   representations carrying additional $U(1)$ charges from certain twisted
   sectors \cite{zurab}.  It is not known whether this loophole can be
exploited in
   realistic (and in particular, chiral) string GUT models.
   Indeed, this seems rather unlikely, given the results of various
   searches and classifications of chiral models that have recently been
   performed.  Thus, the technical results of this paper --- such as those
   of Table~3 --- are likely to continue to remain valid in realistic
   string GUT models, while our more general observations ---
   that free-field string
   constructions pose additional constraints on the allowed matter
   representations in higher-level string models;  that ``effective''
   conformal dimensions can be used to rule out the existence of certain
   large representations that would otherwise be allowed;  that diagonal
   embeddings are very inefficient ways of realizing higher-level gauge
   symmetries in string theory;  and that non-diagonal embeddings
   may have improved phenomenological prospects --- clearly remain valid
   in any case.

   I wish to thank Z. Kakushadze for pointing out the existence of
   the non-chiral $SO(10)$ model.

\bigskip
\bigskip

\bibliographystyle{unsrt}

\end{document}